\documentstyle[11pt,paspconf,epsf]{article}
\def\lesssim{\mathrel{\hbox{\rlap{\hbox{\lower4pt\hbox{$\sim$}}}\hbox{$<$}}}}

\begin{document}

\title{The Ten-Year Photometric Evolution of SN 1987A}
\author{Nicholas B. Suntzeff\altaffilmark{1}}

\affil{Cerro Tololo Inter-American Observatory, National Optical
Astronomy Observatories, Casilla 603, La Serena, Chile}

\altaffiltext{1}{The National Optical Astronomy
Observatories are operated by the Association of Universities for
Research in Astronomy, Inc., under cooperative agreement with the
National Science Foundation}

\begin{abstract}

Ten years of photometric observations of SN 1987A are reviewed. The
optical and near-infrared colors are now declining at $\lesssim 1$
magnitude per 1000 days, which is consistent with both the infrared
``freeze-out'' and the possible energy sources powering the
nebula. The ``uvoir'' bolometric luminosity at ten years is estimated
to be log$_{10}$(L) $\sim 36.1-36.4$ ergs s$^{-1}$. The most recent
photometric measurements are given in Table \ref{t1}. A deep
color-magnitude diagram shows that the young stars are concentrated
within a projected distance of $\sim 25$\arcsec\ (6pc) in a field out
to $\sim3$\arcmin\ from the supernova. It is likely that Stars 2 and 3
which project to within 0.7pc of the supernova are associated with Sk
--69\deg202.

\end{abstract}


\section{Introduction}

With the announcement of a supernova in the Large Magellanic Cloud
(Shelton 1987), many astronomers in the southern hemisphere stopped
the projects they were working on, and began programs of the long-term
monitoring of this unique object.  A tremendous literature has
developed on the observation and theory of the supernova. At the time
of this conference, there are more than 1000 refereed papers listed
under ``SN 1987A'' by the Astrophysics Data System of NASA. In this
short contribution, I would like to summarize the observational work
on the photometry and spectrophotometry of SN 1987A, and discuss the
stellar population surrounding the supernova.

\section{Photometry of SN 1987A}

In the years following the explosion of Sanduleak --69\deg202 on UT
7:36 23 Feb 1987 or JD 2446849.8165 (Bratton {\it et al.} 1988, Hirata
{\it et al.} 1988), all major southern hemisphere observatories
followed the photometric evolution SN 1987A in optical and infrared
colors. In addition, IUE and later HST measured photometric
indices. For ground-based observatories, it was not always easy
finding a telescope {\it small} enough to follow the supernova. The
ground-based observations were also complicated by the fact that the
only detector generally available was a single channel device such as
a photoelectric photometer. A measurement therefore required
photometric weather (not a problem today with CCD detectors and
fainter SNe) or at least scattered clouds so that quick differential
photometry could be done relative to a nearby standard like $\delta$
Doradus. At maximum light on day 84.5, the supernova was $V=2.975$
(Hamuy \& Suntzeff 1990). At this brightness, we had to reduce the
aperture of the 0.4m telescope at CTIO by using the Hartmann mask, and
masking off all but 5 holes, leaving a collecting area of only 26
cm$^2$! Future bright events may leave observatories unprepared as
small telescopes are closed in the face of financial pressures.

\subsection{Atlases of Photometry}

Each observatory published the photometric data in a number of
different papers. Generally the last paper in the series contains
references to the earlier papers. Here I will list only the latest
paper from the series of publications of a particular observatory
group. The number is parenthesis gives the date (since explosion) of
the last observation. In some cases the data are only presented in
graphical form.

\begin{itemize}

\item{\bf AAO/MSSSO:} near-infrared spectrophotometry (day 1114),
Meikle {\it et al.} (1993)

\item {\bf CTIO:} $UBVRI$ photoelectric photometry (day 813), Hamuy \&
Suntzeff (1990); CCD optical photometry (day 1469), Walker \& Suntzeff
(1991); $UBVRIJHK[10][20]$ and ``uvoir'' bolometric luminosities (day
1594), Suntzeff {\it et al.} (1992); optical spectrophotometry (day
805) Phillips {\it et al.} (1990); near-infrared spectrophotometry
(day 1445), Bautista {\it et al.} (1995)

\item{\bf ESO:} Near and mid-infrared photometry and
 spectrophotometry (day 1493), Bouchet \& Danziger (1993), Bouchet
 {\it et al.} (1996); 
Stromgren photometry(day 330), Helt {\it et al.} (1991); 
spectrophotometry (day 314), Hanuschik {\it et al.} (1994); 
Walraven photometry (day 120, some data quoted up to
 day 393 in Pun {\it et al.} 1995), Pel {\it et al.} (1987); 
Geneva photometry (day 917), Burki {\it et al.} (1991)

\item{\bf KAO:} Near and mid-infrared spectrophotometry (day 775),
Wooden {\it et al.}  (1993)

\item{\bf IUE:} Spectrophotometry and  ultraviolet magnitudes (day 1567),
 Pun {\it et al.} (1995);
FES ($V_{IUE}$) light curve (day 760)  Kirshner \& 
 Gilmozzi (1989) (see also Sonneborn 1988)

\item{\bf LCO:} $UBVRI$ photoelectric photometry and early
 photographic photometry (day 156), Shelton(1993a,b), Shelton \&
 Lapasset (1993)

\item{\bf HST:} Ultraviolet magnitudes (day 2431), Pun {\it et al.} (1995)

\item{\bf SAAO:} $UBVRIJHKLM$ single-channel and CCD photometry (day
1770), Caldwell {\it et al.} (1993). Also uvoir bolometric magnitudes.

\item{\bf Astron station:} ultraviolet spectrophotometry (day 394),
Lyubimkov (199)

\end{itemize}

I also list a number of papers which are relevant to the photometric
data: finding chart and star names, Walborn {\it et al.} 1987 and
Walker \& Suntzeff (1990); detailed photometry and spectroscopy of
Stars 2 and 3, Walborn {\it et al.} (1993) ; ``uvoir'' bolometric
luminosities from optical and infrared spectrophotometry, Suntzeff \&
Bouchet (1990), Hamuy {\it et al.} (1990), and Bouchet {\it et al.}
(1991); summary of polarization measurements through day 600, Wang \&
Wheeler (1996); HST spectrophotometry of Star 2 (and a discussion of
age and reddening of this star), Scuderi {\it et al.} (1996).

Although there is a good local photometric sequence near SN 1987A in
the optical (Walker \& Suntzeff 1991, Walborn {\it et al.} 1993), the
photometric observations are by no means easy. The supernova, which is
now at $m \sim 17-20$, is within 3\arcsec\ of Stars 2 and 3 which have
magnitudes of $15 < V < 16$.  Walborn {\it et al.} (1993) show that
Star 3 is a variable Be star with $\delta(V) \sim 0\fm6$ and star 2
shows signs of also being a star Be star. Even with excellent local
standards and image quality, there can be systematic differences as
large as $\delta(m) \sim 0\fm4$ in the photometric data published by
different observatories due to the non-stellar flux distribution of
the supernova in the nebular phase coupled with the different CCD
quantum efficiency/filter combinations (Menzies 1989, Hamuy {\it et
al.}  1990).  The near-infrared photometry is more accurate since the
crowding hot stars are fainter and the image quality is generally
better, but the local standards are only accurate to about 0\fm1
(Suntzeff {\it et al.}  1991, Walborn {\it et al.}  1993).

\subsection{The Light Curve of SN 1987A}

In Table \ref{t1} I list the most recent photometric data for SN
1987A. The HST WFPC2 optical photometry was measured by P.~Challis,
The near-infrared photometry was measured from CTIO 4m telescope
IR imager data.

\begin{table}[h]
\caption{Optical and Near-Infrared Photometry of SN 1987A \label{t1}}
\begin{tabular}{lccccc}
\tableline
\multicolumn{1}{c}{Date} &
\multicolumn{1}{c}{$U$} &
\multicolumn{1}{c}{$B$} &
\multicolumn{1}{c}{$V$} &
\multicolumn{1}{c}{$R$} &
\multicolumn{1}{c}{$I$} \\
\tableline
2927 (3 Mar 95) & ... & 19.68& 19.72& 18.63& 19.07 \\
3267 (6 Feb 96) &   19.85 & 19.93& 19.96& 18.88& 19.37 \\
decline rate\tablenotemark{a}    & 0.89:    & 1.03(13)&0.63(05) &0.82(03)&0.96(02)\\
&&&\\
\tableline
\multicolumn{1}{c}{Date} &&
\multicolumn{1}{c}{$J$} &
\multicolumn{1}{c}{$H$} &
\multicolumn{1}{c}{$K$}  \\
\tableline
3274 (10 Feb 96) &&    17.71& 17.28& 17.14& \\
3558 (20 Nov 96) &&    18.24& 17.63& 17.46& \\
3616 (17 Jan 97) &&    18.01& 17.41& 17.34& \\
3621 (22 Jan 97) &&    18.11& 17.54& 17.22& \\
decline rate\tablenotemark{a}     && 0.75(13) & 0.78(10) & 0.45(11) \\
\tableline
\tableline
\tablenotetext{a}{Decline rate calculated over days 2500-3600 in units
of magnitude (1000 days)$^{-1}$. The error is the standard deviation
of a single observation, in units of 0\fm01.}
\end{tabular}
\end{table}

\begin{figure}[h]
\plotone{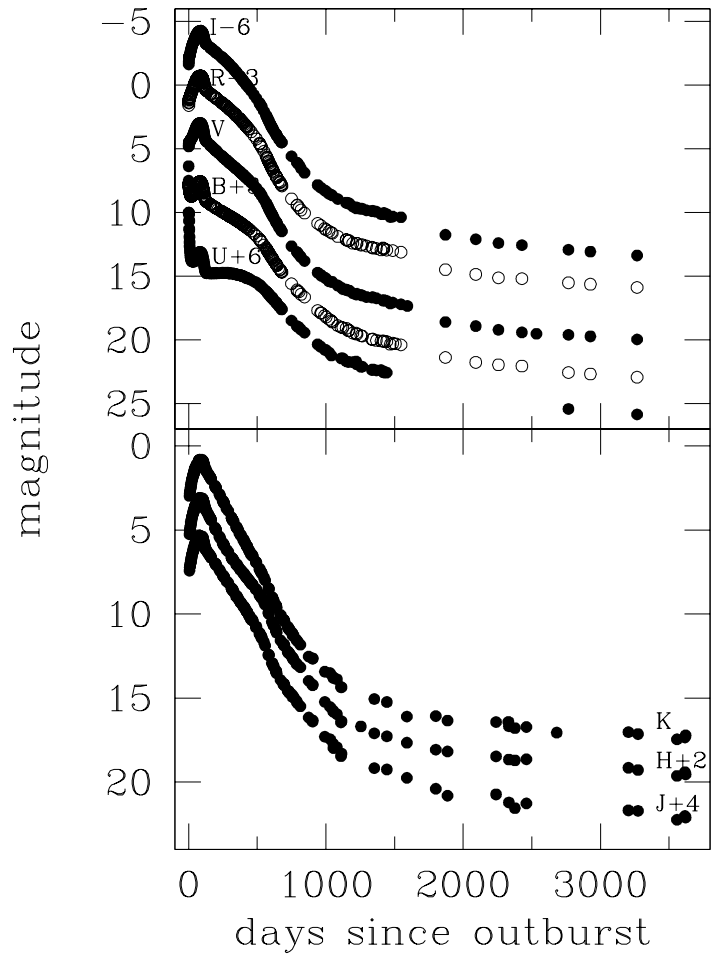}
\caption{$UBVRI$ and $JHK$ photometry of SN 1987A through February
1997. The photometric data have been shifted by (+6,+3,0,-3,-6) for
$UBVRI$ and (+4,+2,0) for $JHK$ \label{f1}}
\end{figure}

The ground-based $JHK$ magnitudes were measured with DAOPHOT (Stetson
1987) psf fitting techniques and do not include flux from the inner
ring. The inner ring adds about 0\fm3  on day 3274. The psf errors
are 0\fm05 (0\fm1 on 3616) and the photometric zero point error is about
0\fm07. The (preliminary) HST magnitudes were measured through a
digital aperture of diameter 0.6\arcsec, with an estimated error of
0\fm15. I also list the decline rate for the data starting around day
2500. In Figure \ref{f1} I plot the $UBVRI$ data from CTIO and HST the
$JHK$ data from CTIO and ESO.

\begin{figure}[h]
\plotfiddle{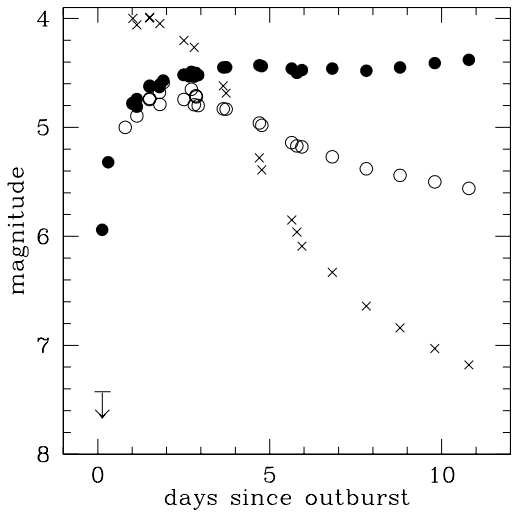}{2.75in}{0}{35}{35}{-125}{-50}
\caption{$UBV$ photometry of SN 1987A through the first 11 days.  The
symbols are: $U$ exes, $B$ open circles, and $V$ filled
circles. \label{f2}}
\end{figure}

The supernova light curve is declining at 1 magnitude per 1000 days or
less, consistent with the possible sources of energy powering the
nebula ($^{44}$Ti, $^{22}$Na, and a pulsar) and the effects of the
infrared ``freeze out'' (Woosley {\it et al.} 1989, Kumagai {\it et
al.} 1991, Fransson \& Kozma 1993).

We cannot directly measure the ``uvoir'' bolometric luminosity at
these late times because we do not have 10-30 micron fluxes. However,
we can use the bolometric corrections to the optical colors $VK$ to
estimate a bolometric magnitude on the assumption that the flux
distribution is indeed frozen. Under this assumption, the uvoir
luminosity on day 3600 is log$_{10}$(L) $\sim 36.1-36.4$ in units of
ergs s$^{-1}$.

From the re-analysis of the early photographic and photoelectric data,
the early-time light curve has been improved (West \& McNaught 1992,
Shelton 1993a, Shelton \& Lapasset 1993). In Figure \ref{f2} I plot
the data from these papers along with the early CTIO, SAAO, and LCO
photoelectric photometry. I also include the photoelectric data
published by Matthews (1987), Moreno \& Walker (1987), and the visual
lower limit of Jones (1987). The Jones limit happened 1.8 hours after
the neutrino detections. The first optical detection was made by
McNaught at 3.1 hours after outburst.

\subsection{The Stellar Population Near SN 1987A}

As summarized by Melnick in this conference, Sk --69\deg202 was
located about 4\arcmin\ from the center of the loose cluster NGC 2044
(LH90) near NGC 2070 (30 Doradus).  The whole region is enveloped in
emission associated with the star formation in the 30 Dor complex. At
the position of Sk --69\deg202 there is no large-scale star formation
activity as in NGC 2044 or NGC 2050 to the south. Walker \& Suntzeff
(1990) however noted a small young association within 30\arcsec\ of
the supernova, which is called KMK80 in the catalog of Kontizas {\it
et al.} (1988). Efrevmov (1991) and Walborn {\it et al.} (1993) have
estimated the age of this association at $12\pm4$Myrs using the CTIO
photometry.  Scuderi {\it et al.} (1996) have estimated an upper limit
(dependent on mass loss) to the age of Star 2 at 13Myrs. This, of
course, is similar to the age estimate for Sk --69\deg202 of 11Myrs by
Arnett {\it et al.} (1989).

\begin{figure}[h]
\plotfiddle{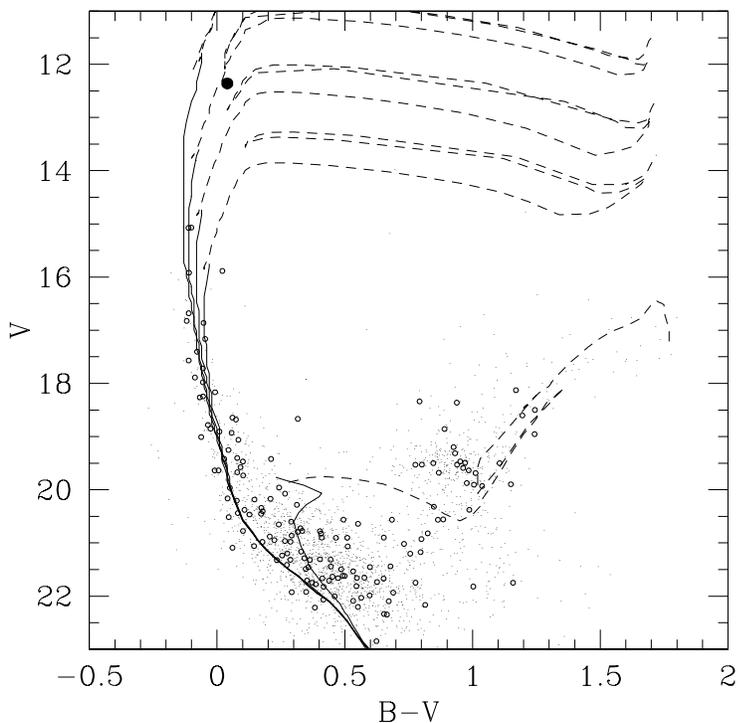}{4in}{0}{50}{50}{-145}{-60}
\caption{$BV$ color-magnitude diagram of the stellar field surrounding
SN 1987A. The stars within 25\arcsec\ of the supernova are marked as
open circles and the stars outside this circle are marked as
points. The whole field extends roughly 2.7\arcmin\ from the
supernova. The position of Sk --69\deg202 is marked as a closed
circle. Isochrones from Bertelli {\it et al.} (1994) for ages 5, 10,
20, 40Myr, and 1Gyr are plotted.  The more rapid evolution on the
giant branch is indicated by dashed lines. The star at $V=16$ which
lies well off the main sequence is Star 3, a known Be variable.
\label{f3}}
\end{figure}

The field population near SN 1987A can be seen in more detail in
Figure \ref{f3}. Here I plot the $BV$ color-magnitude diagram along
with the (Y,Z)=(0.25,0.008) isochrones from Bertelli {\it et al.}
(1994) and the photometry of Sk --69\deg202 from Walborn {\it et al.}
(1987). I have used a reddening of $E(B-V)=0.16$ and a true distance
modulus of 18.50. These data were taken in January and March 1996 at
the CTIO 4m telescope CFCCD system by Mark Phillips. A total of seven
frames in each color were used to calculate the magnitudes. The large
scatter about the main sequence down to $V=20$ is not due to
photometric errors, but due to variable reddening (Walborn {\it et
al.} 1993).

Figure \ref{f3} shows that the young field population within
25\arcsec\ of the supernova is consistent with a population of
$\sim15$Myr, although this value critically depends on the accuracy of
the theoretical $T_{eff}$ to $(B-V)$ conversion, the reddening, and
the small number of hot stars. There is also a dominant older field
population with a prominent turnoff at $V=20.4$ which corresponds to a
1.5Gyr population.

These two populations are not well mixed however. Figure \ref{f3}
shows that {\it the young field population is strongly concentrated
towards the position of SN 1987A}. This can be shown with simple
statistics. Divide the field by a circle with radius of 25\arcsec\
(projected distance of $\sim 6$pc) around SN 1987A and count the
number of hot stars, defined here simply as any star with $(B-V) <
0.5$. Form the ratio of the number of stars with $V < 18$ and $18 < V
< 20$. The ratios for inside and outside the circle are 14/23 and
20/178 or 61\% versus 17\%. Furthermore, of the eight brightest hot
stars in this region ($V<16$), four are inside the circle (stars
2,3,10,30), and two are inside 3\arcsec! Given such statistics, it is
extremely likely that Sk --69\deg202 was part of a loose association
including KHK80 with an age of $\sim15$Myr, and that Stars 2 and 3,
which are less than 3\arcsec away, are physically associated with
progenitor of the supernova. However, because there is no obvious
interaction of the outer rings of the supernova with Stars 2 and 3, it
is unlikely that the supernova lies in precisely the same plane on the
sky as the two crowding stars.


\acknowledgments

This paper is the summary of the work of many astronomers who have
collaborated to bring this photometry to the astronomical community.
I would like to acknowledge the help of 
Patrice Bouchet,
Peter Challis,
Darren DePoy,
Jay Elias,
Richard Elston,
Christian Gouiffes,
Mario Hamuy,
Jason Pun,
Bob Kirshner,
Jaymie Matthews,
Mark Phillips,
S. Elizabeth Turner,
Alistair Walker,
and
Robert Williams.
The help of Jason Spyromilio with the pronunciation of ``uvoir'' is
always appreciated.  I also gratefully acknowledge the continuing
support for research on SN 1987A through the HST grants
``SINS: Supernova INtesive Study, '' Robert Kirshner, PI, and, of
course, NOAO.

\end{document}